\documentclass[aps,prb,twocolumn]{revtex4}
\usepackage{psfig}
\usepackage{amssymb}

\begin{document}
\title{Huygens description of resonance phenomena in subwavelength hole arrays}

\author{C. Genet, M.P. van Exter, and J.P. Woerdman}

\affiliation{Huygens Laboratory, Leiden University, P.O. Box 9504,
2300 RA Leiden, The Netherlands}

\begin{abstract}
We develop a point-scattering approach to the plane-wave optical
transmission of subwavelength metal hole arrays. We present a {\it real} space description instead of the more
conventional {\it reciprocal} space description; this naturally produces interfering resonant features in the
transmission spectra and makes explicit the tensorial properties of the transmission matrix. We give transmission
spectra simulations for both square and hexagonal arrays; these can be evaluated at arbitrary angles and
polarizations.
\end{abstract}

\maketitle

\begin{figure}[tbh]
\centerline{\psfig{figure=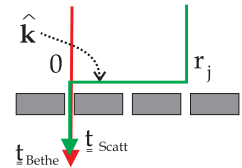,width=3cm}} \caption{Schematic representation of the scattering process,
including the direct scattering contribution defined at a chosen origin of the lattice.}  \label{scheme}
\end{figure}

\begin{figure}[tbh] \centerline{\psfig{figure=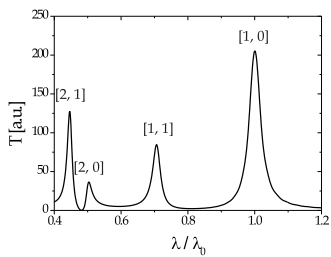,width=5cm}}
\caption{Transmission coefficient of a square array at normal incidence, as a function of normalized input
wavelength.} \label{fresnelzone}
\end{figure}

\begin{figure}[tbh]
\centerline{\psfig{figure=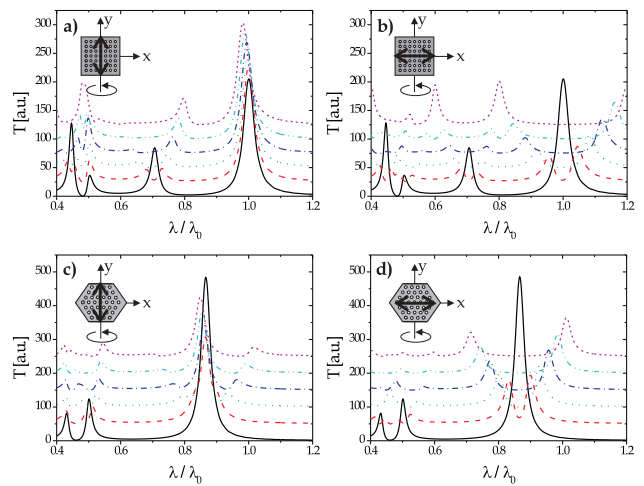,width=9cm}} \caption{Transmission spectra for a square array (first row) for
a)TE and b)TM polarizations and for a hexagonal array (second row) for c)TE and d)TM polarizations. These spectra
have been calculated as a function of normalized input wavelength and angle of incidence when the tilt is
performed along the $y-$axis - see insets. For clarity, we show a globally shifted curve for each angular
increment of $\sim 2.3^{\rm o}$.} \label{spectrum}
\end{figure}

\indent Experiments have revealed the crucial role played by surface wave excitations (often called surface
plasmons) in the case of extraordinary transmission features of nanoperforated metallic
films\cite{EbbesenNature1998}. This experimental work has generated an important theoretical literature which can
be grossly separated into two parts. A first category is based on {\it ab initio} fully numerical simulations of
the scattering amplitudes of these nanoarrays; in this case, the interpretation of simulated spectra is often
difficult\cite{PopovPRB2000,MartinMorenoPRL2001,EnochJOptA2002,SarrazinPRB2003}. A second category is accordingly
devoted to the search for physical understanding of the
phenomenon\cite{GhaemiPRB1998,VigoureuxOptComm2001,DarmayanPRB2003}. The main criticism that can be addressed to
these interpretative papers is that they rely on the {\it a priori} definition of resonances as surface modes at a
{\it smooth} interface (i.e. no holes) which are coupled to the incident field via momentum matching by the hole array.\\
\indent The aim of this paper is to fill the intermediate gap with a simple model which allows both a clear
physical description and straightforward simulations of transmission spectra. The main attractive features of our
model are related to the fact that it is based on a {\it real} space (instead of a {\it reciprocal} space) surface
wave scattering analysis. We stress that our work naturally produces transmission resonances and that it does so
without invoking {\it a priori} wavevector matching between surface modes on a {\it smooth} interface and grating
momenta, i.e. reciprocal lattice vectors. Most importantly, it reveals interference effects between neighbouring
resonances, contrary to the usual reciprocal space (i.e. Fourier) description of wavevector matching. Our
description makes explicit, among others, the polarization dependence and the tensorial properties of the
transmission matrix. It describes the influence of the incident angle tuning on transmission spectra and,
eventually, it naturally relates corresponding band structures to symmetries of the {\it reciprocal} lattice of
the array. All these aspects are, for the sake of demonstration, best clarified in a rather simplified framework,
as we will discuss below. This implies that we do not aim at quantitative agreement with experimental transmission
spectra. Nevertheless, we feel that our model yields original physical insights into the dynamics of transmission
through nanohole arrays.\\
\indent Our approach is rooted in the role played by surface modes on the transmission process. The transmission
process of a plane wave through sub-wavelength metallic hole arrays will be cast into the context of a
Huygens-type principle, where the array is discretized as a lattice of holes acting as point-scatterers that
scatter the incident radiation coherently into two-dimensional secondary wavelets. Specifically, the array is
contained in the $(x,y)$ plane of a cartesian $(x,y,z)$ frame and is illuminated by a paraxial plane wave of
wavevector ${\bf k}_{\rm in}$ and polarization vector $ \hat{{\bf u}}_{\rm in}$. This wave is described by
far-field angles ${\bf k}_{\rm in} / \left|{\bf k}_{\rm in}\right|\sim (\theta_{x},\theta_{y},1)$ and two electric
field components ${\bf E}\sim ({\bf E}_{x},{\bf E}_{y})$, decomposed into the basis of the TE and TM
polarizations. Retaining only the $0^{\rm th}$ diffraction order, the transmission can be formulated as ${\bf
E}_{\rm out}[\lambda,\theta ]=\underline{\underline{{\bf t}}}\cdot {\bf E}_{\rm in}[\lambda,\theta ]$ with a
$2\times 2$ transmission matrix $\underline{\underline{{\bf t}}}[\lambda,\theta ]$, $\lambda$ being the wavelength
and
$\theta=(\theta_{x},\theta_{y})$ the far field angles of the incoming and outgoing field.\\
\indent As schematically shown in fig.(\ref{scheme}), we will distinguish two contributions to this transmission
matrix \cite{SarrazinPRB2003,GenetOptComm2003}. A first one, $\underline{\underline{{\bf t}}}_{\rm Bethe}$,
wavelength dependent and proportional to the identity matrix, corresponds to a transmission of the incoming field
directly through the holes, i.e. to a Bethe-type diffraction regime \cite{BethePR1944}. A second contribution,
$\underline{\underline{{\bf t}}}_{\rm Scatt}$, corresponds to the resonant part of the transmission matrix, on
which we focus hereafter. Fig.(\ref{scheme}) describes this resonant transmission process as a three-step process:
(i) the incident plane wave is converted into a surface wave at a given point scatterer, (ii) the surface wave
propagates on the surface of the array and (iii) is eventually re-emitted as a plane wave through the array.
Then, in the spirit of Fresnel diffraction, the global resonant matrix $\underline{\underline{{\bf t}}}_{\rm
Scatt}$, evaluated at the center $({\bf 0})$ of the array, is based on the {\it near-field} of the array, as the
coherent resummation over all the secondary surface wavelets emitted by each of the holes, acting as independent
uncoupled holes with elementary scattering matrices $\underline{\underline{{\bf \tau}}}$
\begin{eqnarray}
\underline{\underline{{\bf t}}}_{\rm Scatt}({\bf 0})=\sum_{ \{{\bf r}_{j}\} }\underline{\underline{{\bf
\tau}}}({\bf 0},{\bf r}_{j}) \label{globalscatt}
\end{eqnarray}
The summation is defined on the lattice distribution of holes $ \{{\bf r}_{j}\}$. The singular term $\{{\bf
r}_{j}={\bf 0} \}$ is related to the direct transmission channel of $\underline{\underline{{\bf t}}}_{\rm Bethe}$
and is therefore naturally excluded from this resonant part. As a periodic lattice, the array is coordinated by
its two primitive vectors $({\bf a}_{1},{\bf a}_{2})$. We restrict this paper to $|{\bf a}_{1}|=|{\bf
a}_{2}|=a_{0}$ square and hexagonal lattices, as these symmetries are most popular
experimentally\cite{MurrayPRB2004,AltewischerPRLsubmitted}. The position of each scattering hole on each of these
lattices is defined as ${\bf r}_{j}=n{\bf a}_{1}+m{\bf a}_{2}$ with $(n,m)$ integers. The summation on
Eq.(\ref{globalscatt}) therefore corresponds to a double $(n,m)\neq(0,0)$ summation over the lattice. \\
\indent In this work, we calculate the $\underline{\underline{{\bf \tau}}}$ matrix from the dynamics of a
scattered surface wave characterized by a complex transverse wavevector $| \hat{{\bf k}}|=(\eta_{1}+i\eta_{2}
)2\pi / \lambda  $. On a {\it smooth} metal-dielectric interface (i.e. no holes), with dielectric functions
$\varepsilon_{1} , \varepsilon_{2}$ respectively, theory predicts\cite{RaetherBook}
\begin{eqnarray}
\eta_{1}+i\eta_{2}=\sqrt{\frac{\varepsilon_{1}\varepsilon_{2}}{\varepsilon_{1}+\varepsilon_{2}}}
\end{eqnarray}
Note that this expression holds for a metal-dielectric interface, from the visible to the microwave domain, but
not for a dielectric-dielectric interface. In practice, the observed transmission resonances in metallic nanohole
arrays are much broader than and red-shifted by typically a few percent from the mode dispersion on a smooth
interface\cite{GhaemiPRB1998,SalomonPRL2001,KrishnanOptComm2001}. These discrepancies are most likely related to
direct Bethe-type transmission channel and to radiative losses of the surface waves when they scatter on the holes
of the actual structure.
In our simulations, we use realistic values for $\eta_{1}$ and $\eta_{2}$ (see below).\\
\indent In our model, the polarization of each surface wave is taken along its propagation direction with a
unitary polarization vector $\hat{{\bf u}}_{j}={\bf r}_{j} / |{\bf r}_{j} |$. Surface polarization plays a crucial
role, as it determines both the incoupling efficiency $\hat{{\bf u}}_{j}\cdot\hat{{\bf u}}_{\rm in}$ between the
free-space incident photons and the excited surface wave, as being proportional to their electric field overlap,
and the polarization of the emitted radiation. The incoupling factor $\hat{{\bf u}}_{j}\cdot\hat{{\bf u}}_{\rm
in}=\cos\varphi$ corresponds to a two-dimensional dipole radiation pattern for the surface wave emitted at the
hole; this pattern has been experimentally observed \cite{HechtPRL1996}. The full polarization behaviour is
contained in the tensorial $\hat{{\bf u}}_{j}\otimes\hat{{\bf u}}_{j}$ nature of the elementary point-scattering
matrix ($\otimes$ denotes a tensorial product).\\
\indent We assume that the elementary scattering matrix $\underline{\underline{{\bf \tau}}}$ is spherical in the
far-field so that it reads as
\begin{eqnarray}
\underline{\underline{{\bf \tau}}}({\bf 0},{\bf r}_{j})=f( | \hat{{\bf k}} | )\frac{e^{i|\hat{{\bf k}}||{\bf
r}_{j}|}}{\sqrt{|{\bf r}_{j}|}}e^{i{\bf k}_{\rm in}.{\bf r}_{j}}\hat{{\bf u}}_{j}\otimes\hat{{\bf u}}_{j}
\label{elemscatt}
\end{eqnarray}
The scattering amplitude is $f( | \hat{{\bf k}} | )=s( | \hat{{\bf k}} | )e^{-i\pi / 4}\sqrt{ {\rm Re}( | \hat{\bf
k} | ) / 2\pi }$ in the far-field approximation $|\hat{{\bf k}}||{\bf r}_{j}|\gg 1$ of the Huygens phase, being
certainly satisfied for $\lambda\ll 2\pi a_{0}\eta_{1}$. For simplicity, we neglect the frequency dependence of
the shape factor $s( | \hat{{\bf k}} | )$ and, in the point-scattering limit, replace it by a constant $s$. This
limits the discussion to spatial symmetries of the lattice. Point-group issues matter when specific shapes of the
scatterers are introduced. Then, the scattering amplitude has to include a true $s(\hat{{\bf k}} )$ shape factor.
The coherence of the surface scattering is insured by the fixed phase relation $e^{i{\bf k}_{\rm in}.{\bf r}_{j}}$
with the incident field ${\bf k}_{\rm in}$. Eventually, with an input plane wave ${\bf E}_{\rm in}={\rm E}_{\rm
in}\hat{{\bf u}}_{\rm in}$ and no polarization analysis in transmission, we evaluate the resonant part of the
intensity transmission coefficient ${\rm T}= | \underline{\underline{{\bf t}}}_{\rm Scatt} \cdot \hat{{\bf
u}}_{\rm in} | ^{2} $. \\
\indent In our simulations, we normalize the wavelength by $\lambda_{0}=a_{0}\eta_{1}$ and the transmission matrix
by the constant scattering amplitude factor $s$. Internal damping and radiative losses of a surface wave
propagating on the lattice are quantified with a single effective parameter $\eta=\eta_{2} / \eta_{1}$. This
$\eta$ determines the convergence speed of the scattering summation.  For simplicity, we neglect the frequency
dependence of both $\eta_{1}$ and $\eta_{2}$ and choose $\eta\sim 0.02$ which correspond to a constant mean free
path of $a_{0}/ (4\pi \eta)\sim 3$ $\mu$m for $a_{0}=0.7$ $\mu$m \cite{AltewisherNature2002}. Convergence is then
easily reached with a lattice of $80\times 80$ points. If the damping is lower, a larger lattice should be chosen.
The frequency dependence of $\underline{\underline{{\bf t}}}_{\rm Scatt}$ is determined by running simulations for
$800$ wavelengths positioned on a regular grid ranging from $\lambda / \lambda_{0}=0.4$ to $\lambda /
\lambda_{0}=1.2$. This interval is very large (factor of $3$) but is nevertheless consistent with our model,
avoiding both shorter wavelength where the direct transmission channel takes over the resonant one, and longer
wavelength for which the far-field approximation might break down. These nominal parameters given, the evaluation
of an element of the transmission matrix,
at a given incidence angle, takes about $5$ seconds for the whole spectrum.\\
\indent As a first working example of our method, we present in fig.(\ref{fresnelzone}) a simulated spectrum of
the transmission coefficient of a square array over the wavelength range mentioned above, under plane-wave
illumination at normal incidence $\theta=0$. In this case, both TE and TM polarizations are degenerate and the
transmission matrix is simply proportional to the identity matrix. Our model immediately shows the resonant
behaviour of the transmission spectrum. We stress again that we make here no use of the usual momentum
conservation argument specifying resonances in terms of reciprocal lattice; the resonances arise naturally from
the summation
of Eq.(\ref{globalscatt}). \\
\indent The position and the relative strength of the peaks shown in the spectrum of fig.(\ref{fresnelzone}) can
be understood from a Fresnel-zone type argument \cite{BornWolf1975}. A resonance builds up each time the Huygens
phase satisfies an in-phase condition, related to the definition of the reciprocal lattice. Resonances are thus
expected at approximately the same positions predicted by the usual {\it ad hoc} dispersion relation for surface
waves on periodic arrays, that is at
\begin{eqnarray}
\frac{\lambda_{\rm res}}{ \lambda_{0}}= \frac{1 }{\sqrt{n^2+m^2}}
\end{eqnarray}
for a square array and $\lambda_{\rm res} / \lambda_{0}= \sqrt{3} / 2\sqrt{n^2+nm+m^2}$ for a hexagonal
array\cite{GhaemiPRB1998}. We adopt hereafter the convention of indexing a surface mode as $[n,m]$. The strength
of the peaks scales with the corresponding inverse distance $1 / \sqrt{n^2+m^2}$. The increase in peak
transmission for the $[2,1]$ mode as compared to the $[2,0]$ mode results from an increase in degeneracy from
$4-$fold to $8-$fold, leading to a factor $2$ increase in amplitude and a factor $4$ in intensity. The spectrum in
fig.(\ref{fresnelzone}) is shown in arbitrary units
(a.u.) where, from Eq.(\ref{globalscatt}), one hole at one lattice spacing contributes one arbitrary unit.\\
\indent We note that the two-dimensional summation on a discrete two-dimensional lattice (see
Eq.(\ref{globalscatt})) does not provide all in-phase conditions. For the square array, for instance, the in-phase
conditions derived from Eq.(\ref{globalscatt}) read as
\begin{eqnarray}
2\pi\sqrt{n^2+m^2}\frac{\lambda_{0} }{\lambda_{\rm res}} =  2\pi p
\end{eqnarray}
with $p$ a given positive integer. However, $n^2+m^2$ can not provide all such $p$ integers. For this reason,
there is no resonance between the $[1,1]$ and $[2,0]$ modes of a square array, though such an in-phase condition
could in principle be satisfied in between. This lack of resonances corresponds to a modification of the simple
Fresnel-zone argument: with a two-dimensional lattice, the only
multiplicities corresponding to resonant conditions are those matching the discretization of the lattice.\\
\indent This picture is useful to understand the positions and strengths of the resonances. But we would like to
stress that our {\it real} space-based model goes beyond that, producing interferences between surface modes that
have a strong influence on the transmission spectrum. Peaks present asymmetries due to the interference of
overlapping tails of adjacent resonances.
These tails can even interfere quite destructively as in between modes $[2,1]$ and $[2,0]$ (see
fig.(\ref{fresnelzone})). The interferences between the modal amplitudes make the actual position of the resonance
$\lambda_{\rm
res} / \lambda_{0}$ slightly different from the ``pure'' location $1 / \sqrt{n^2+m^2}$.\\
\indent We have also investigated non-normal incidences that correspond to tilts along the $y$ axis, for both
square and hexagonal arrays (oriented with one lattice vector in the $x-$direction). Rotation angles ranging from
$0^{\rm o}$ to $\sim 11.5^{\rm o}$ in steps of $\sim 2.3^{\rm o}$ were studied. TE and TM polarizations, now
distinct, are displayed in separate figures. A central result of this paper, shown in figs.(\ref{spectrum}), is
that our model immediately
and naturally produces spectral band structures.\\
\indent For a square array, illuminated with a TE polarized wave, the $[1,0]$ resonance remains practically
stationary, whereas it splits into a doublet for a TM polarization as the angle of incidence is increased. For the
$[1,1]$ mode, one observes a doublet splitting for both polarization. For a hexagonal array, the principal
resonance splits into a triplet with a central peak for TE polarization, whereas it evolves into a doublet for TM
polarization. Split resonances will eventually overlap, through their progressive shifts. This can lead to a
destructive interference, as seen at an $\sim 11.5^{\rm o}$ incidence angle on fig.(\ref{spectrum}, a)) between
the $[1,1]$ and $[2,0]$ resonances for TE polarization or to a constructive interference, as seen on
fig.(\ref{spectrum}, b)) between the $[1,0]$ and $[1,1]$ modes for TM polarization. One should also note that for
both square and hexagonal arrays, central peaks are in fact only approximately stationary as the tilting angle is
varied. With an increasing angle of incidence, a positive transverse component of the
wavevector ${\bf k}_{\rm in}$ emerges and leads to a slight blue shift of central peaks.\\
\indent It is interesting to compare these simulations, and in particular the splitting of resonances, with what
could be expected from rather intuitive arguments. Intuitively, a nanohole array is simply characterized by the
spatial symmetries of its {\it direct} lattice. The direction of propagation of a given surface mode will be
naturally associated to a particular axis of this direct lattice, with an incoupling efficiency argument as
presented above and based on the projection factor between such a propagation axis and the incident polarization.
Thus, one forbids modes which propagate along directions perpendicular to the incident polarization to be excited.
Within this intuitive frame, one predicts splitting of resonances by merely performing symmetry operations on the
{\it direct} lattice as the array is tilted. It turns out however that following this line of
reasoning leads to wrong predictions. \\
\indent On the contrary, it is important to realize that the band structures of figs.(\ref{spectrum}) can only be
fully inferred if one realizes that Eq.(\ref{globalscatt}), starting indeed from point-scatterers on the {\it
direct} lattice of the array, provides resulting waves that are directional and that propagate along axes of the
{\it reciprocal} lattice rather than the {\it direct} one. Therefore, these band structures are only consistent
with the symmetry arguments performed on the {\it reciprocal} lattice\cite{AltewischerJOSAB2003}, together with a
polarization ``selection-rule'' killing surface
waves that are allowed by symmetry to propagate on {\it reciprocal} lattice axes perpendicular to the incident
polarization. \\
\indent The disagreement with the intuitive approach mentioned above is not present for a square array since in
this case, the direct lattice and the reciprocal lattice are identical. It is however easily visible for a
hexagonal array for which direct and reciprocal lattices do not coincide. In this case, if one concentrates the
symmetry arguments on the direct lattice, the expected splitting of resonances will not agree with simulations.
For instance, symmetry arguments applied to a direct hexagonal lattice tilted along its $y-$axis lead to the
prediction of quadruplet splitting with no stationary mode for the principal resonance, for both polarizations.
The same argument applied to the {\it reciprocal} lattice, together with accounting for the projection factor
between lattice axes and polarization, leads to a triplet, respectively a doublet, for TE, respectively TM,
polarization; this is indeed
observed in fig.(\ref{spectrum}, c)) and fig.(\ref{spectrum}, d)).\\
\indent In conclusion, we have presented a simple and straightforward model based on a Huygens-type principle.
This work provides physical insight into the surface-plasmon-assisted transmission process through metal nanohole
arrays, emphasizing symmetry and tensorial properties of the transmission amplitude. It yields band structures
consistent with directional collective surface excitations propagating on the {\it reciprocal} lattice of the
nanohole
array and does not rely on any {\it ad hoc} momentum matching argument.\\
\indent In order to stress the core characteristics of our Huygens description, we have used a simplified model
that addresses the resonant contribution $\underline{\underline{{\bf t}}}_{\rm Scatt}$ to the transmission through
a single interface only and that does not account for the direct transmission channel contribution
$\underline{\underline{{\bf t}}}_{\rm Bethe}$. Our point-scattering approach amounts to considering the holes in
the metallic film to be in the far subwavelength limit, i.e. basically in the limit of holes of zero diameter. To
quantitatively compare with experimental data, the following extensions are needed: both the dielectric constant
of the metal and the scattering amplitude $s(\hat{{\bf k}})$ require a frequency dependence. Furthermore, related
in particular to hole size effects, this scattering amplitude should be specified and the direct contribution
$\underline{\underline{{\bf t}}}_{\rm Bethe}$ to the transmission should be included, providing resonance line
shapes with red-shifts and red-tails \cite{GenetOptComm2003}. \\
\indent Of course, a qualitative agreement between our simulations and angle dependent transmission measurements
should be easier to obtain, in particular as far as splittings of resonances are concerned. But discrepancies are
already found at this less ambitious level when referring to experiments performed with standardly designed
nanohole arrays\cite{Ebbesen Nature1998,GhaemiPRB1998}. In fact, such nanohole arrays correspond to optical
systems with two-interfaces, possibly identical in the case of free-standing films. There, angle dependent
couplings between surface modes defined on each interface induce perturbations too strong to allow even a
qualitative matching with the mere single-interface band structure provided by our model. Nevertheless, if one
could design an effective single-interface nanoperforated film, such a qualitative check should be possible. A
metallic film, with very small holes, deposited on a given dielectric substrate with a thick titanium (Ti) bonding
layer is likely to be a relevant candidate\cite{GenetOptComm2003}. Strong absorbtion in this Ti layer prevents
indeed any surface mode from being excited on this metal-dielectric interface, thus keeping only a single
air-metal interface into play. In this framework therefore, it will be interesting to confront experimental
transmission spectra with the Huygens description formulated in our paper.

\end{document}